# Formation and Control of Twin Domains in the Pyrochlore Oxide $Cd_2Re_2O_7$

Yasuhito Matsubayashi, Daigorou Hirai, Masashi Tokunaga, and Zenji Hiroi[*]

*Institute for Solid State Physics, University of Tokyo, Kashiwa, Chiba 277-8581, Japan*

The successive phase transitions of the pyrochlore oxide $Cd_2Re_2O_7$ are studied by polarizing microscopy and magnetic susceptibility measurements. The formation of twin domains is visualized in the polarizing images of a pristine (111) crystal surface upon cooling through the cubic-to-tetragonal transition at $T_{s1} \sim$ 200 K. Moreover, a dramatic change in the twinning pattern is observed at $T_{s2} \sim$ 120 K, which suggests that the tetragonal $c$ axis flips as the strain changes its direction at the tetragonal-to-tetragonal transition. Magnetic susceptibility measurements reveal significant domain alignment upon cooling across $T_{s1}$ and $T_{s2}$ in a magnetic field of 7 T, which are due to ~10% anisotropy in the magnetic susceptibility for the low-temperature phases. Interestingly, the anisotropy is reversed at $T_{s2}$: $\chi_c < \chi_a$ above $T_{s2}$ and vice versa below $T_{s2}$.

## 1. Introduction

The pyrochlore oxide $Cd_2Re_2O_7$ has attracted great attention as a candidate spin–orbit-coupled metal (SOCM).[1,2] An SOCM is a metal with a centrosymmetric crystal structure and a large spin–orbit coupling. It possesses a specific Fermi liquid instability that causes spontaneous inversion-symmetry breaking and leads to either multipolar, ferroelectric, or gyrotropic order.[1] The successive phase transitions that $Cd_2Re_2O_7$ exhibit upon cooling have been considered as a good example of multipolar ordering in an SOCM.[3-5]

The two transitions of $Cd_2Re_2O_7$ occur at $T_{s1} \sim$ 200 K and $T_{s2} \sim$ 120 K; with decreasing temperature, the three phases are named phases I, II, and III.[2,6] These transitions are accompanied by characteristic structural changes from centrosymmetric cubic $Fd\bar{3}m$ for phase I to non-centrosymmetric tetragonal $I\bar{4}m2$ for phase II and to another non-centrosymmetric tetragonal $I4_122$ for phase III; the first transition is of the second order and the second one of the first order.[6] In phase I, the $Re^{5+}$ ions with the $5d^2$ electron configuration form a pyrochlore lattice made of regular tetrahedra, while it is slightly distorted in phases II and III. Since the tetragonal distortions are very small, at most 0.05%,[7] the electronic instability of the SOCM may be the main driving force for these transitions, possibly to multipolar orders.[1,3]

In general, a symmetry-lowering transition accompanied by a change in the unit cell shape, i.e., strain, generates intergrown domains with different orientations, as in the case of martensitic transitions[8] or ferroelastic transitions.[9] For example, when a cubic crystal is transformed to a tetragonal crystal, losing its threefold axis, three types of domains are generated depending on the direction of the fourfold $c$ axis. At this transition, twinning often takes place between two types of domains separated by a 90º domain wall in order to minimize the global strain energy over the crystal. Such twins are called deformation twins and are typically observed in martensitic transitions.[8,10] Each domain appears as a lamella or a lens with a specific crystal plane at the interface with another domain. The interface between domains is called the invariant plane, which is unaffected by the structural transformation. The two types of domains stack alternately with the direction of strain changed by 90º, resulting in microstructures typically of the micrometer scale. The size of the domains is determined by the balance between the strain and interface energies. The twins cannot be removed without applying an effective external force.

The cubic-to-tetragonal transition at $T_{s1}$ in $Cd_2Re_2O_7$ results in deformation twins. Harter et al. reported the formation of lamellar domains of ~100 μm thickness at 150 K in their polarizing microscope images without identifying their orientations.[3] Moreover, they observed two regions in one twin domain, which they suggested to be domains of an odd-parity nematic order on the basis of second-harmonic optical anisotropy measurements. However, the detail of the domain formation and, in addition, what happens at $T_{s2}$ remain unknown.

In order to elucidate the nature of the possible multipolar orders of $Cd_2Re_2O_7$, it is crucial to study the electronic anisotropy of the tetragonal phases. For this, a crystal with a single-domain structure has to be examined in experiments. Thus, domain formation and its control by external forces should be investigated. In actual crystals, however, the distribution and mobility of domain walls are often influenced by extrinsic effects such as crystalline defects. Thus, to carry out reliable and reproducible experiments, a high-quality crystal with fewer defects is required. We have succeeded in obtaining high-quality crystals of $Cd_2Re_2O_7$ by improving the crystal growth technique.[2,11] One appropriate measure for the crystalline quality of metallic compounds is the residual resistivity ratio (RRR), which is defined as the residual resistivity divided by the resistivity at room temperature. The crystals used in the previous studies by many groups had RRR values of at most 40, while our recent crystals had larger RRRs of 100–300. These crystals enabled us to observe de Haas–van Alphen oscillations to capture the spin-split Fermi surfaces.[11]

Using high-quality $Cd_2Re_2O_7$ crystals in the present study, we carried out polarizing microscopy observations at low temperatures down to 8 K. We clearly demonstrated the formation of twin domains with a thickness of 10 μm order below $T_{s1}$ and a dramatic change in the domain pattern at $T_{s2}$. We also examined the effects of applying a magnetic field on the magnetic susceptibility at the two transitions, which illustrates the possibility of domain control utilizing the magnetic anisotropy of ~10% in both tetragonal phases. A few possible routes to attaining a single-domain crystal will be discussed.

## 2. Experimental Procedure

Crystals of $Cd_2Re_2O_7$ were grown by the chemical transport method, the detail of which will be reported elesewhere.[12] Octahedral crystals with a size of mm order and RRR = 5–150 were used for measurements. Polarizing microscopy observations were carried out using an optical microscope (Olympus, BXFM) in the reflection mode. Images were recorded using a monochrome CCD camera (ST-402ME, SBIG). A pristine (111) crystal surface was observed in a crossed-Nicols setup with a polarizer along [1–10] and with an analyzer along [11–2] in most observations; if not mentioned specifically, the direction is based on the cubic structure of phase I. A crystal was attached on a sapphire plate with varnish and cooled slowly from room temperature to 8 K in vacuum.

In general, an isotropic crystal appears dark in an image in the crossed-Nicols setup with the polarizer orthogonal to the analyzer. An anisotropic crystal appears bright, as the polarization of incident light can be rotated by the anisotropy in the refractive index of the crystal, allowing reflected light to pass through the analyzer. This interference color changes with the direction of the optical axis of the crystal. In the two low-temperature tetragonal phases of $Cd_2Re_2O_7$, which are optically uniaxial, the interference color changes with the direction of the $c$ axis: extinction occurs for the $c$ axis along either of the polarizer or the analyzer, and a crystal appears brightest for 45º rotation. In our experimental setup, however, we could not determine the direction of the $c$ axis in one domain uniquely from the interference color because the polarization of light was also rotated at the mirrors. Instead, we determined the $c$ axis from the orientation of twin interfaces.

We also attempted to observe a (001) surface that was obtained by polishing an octahedral crystal. However, no interference color was detected. The surface may have been degraded by polishing, which has actually been observed by various measurements in previous studies.[2] We are currently attempting to obtain a crystal with a (001) facet via fine tuning of the crystal growth conditions.

Magnetic susceptibility was measured in a magnetic property measurement system 3 (Quantum Design, MPMS3). By using the vibrating sample magnetometer mode, a high sensitivity of $8 \times 10^{-8}$ emu at 7 T was achieved, which enabled us to obtain reliable data on the anisotropy in the magnetic susceptibility from a single crystal; the magnetization had previously been calibrated with a palladium reference sample. A 2-mm-size crystal of 35.23 mg weight was attached to a semicylinder of quartz with vanish and slowly cooled to 2 K. In actual measurements, data were slightly scattered among different temperature runs even for the same crystal, which may have depended on the lateral sample position inside the pick-up coil and also on the orientation of the crystal with a nonspherical shape. In order to remove the scattering, all data were adjusted so as to match the data measured at 200 K (at which there is no anisotropy) for $H \parallel [111]$ by multiplying by appropriate factors, which were at most 4%. Thus, the absolute values may have some ambiguity, while the relative values should be reliable.

## 3. Results
### 3.1 Observations of tetragonal domains

Since the $T_{s1}$ transition of $Cd_2Re_2O_7$ is a transformation from face-centered cubic to body-centered tetragonal, twin domains with invariant planes of {110} should occur.[9,10] Thus, six types of deformation twins with {110} invariant planes are generated. Figure 1 shows three polarizing microscope images taken at 10 K from different crystals with RRR = 150, 30, and 5, and Fig. 2 shows an image at 8.8 K taken from another crystal with RRR = 60. The interference color strongly depends on the crystal: the crystal in Fig. 2 has a larger contrast variation than the others in their original photographs; the images in Fig. 1 have been modified to enhance the contrast for clarity, while that in Fig. 2 has not been changed.

By comparing the three crystals in Fig. 1, it is observed that as the RRR decreases, the contrast tends to become blurred, the domain width is reduced, and the interface between the domains becomes irregular. Thus, there is a tendency that the domain formation is disturbed with decreasing RRR; crystalline defects or inclusions that cause carrier scattering to decrease the RRR probably promote the formation of domains and pin the domain walls. Nevertheless, note that the RRR may not be the only parameter affecting domain formation, and a complex mechanism may be involved in the domain formation.

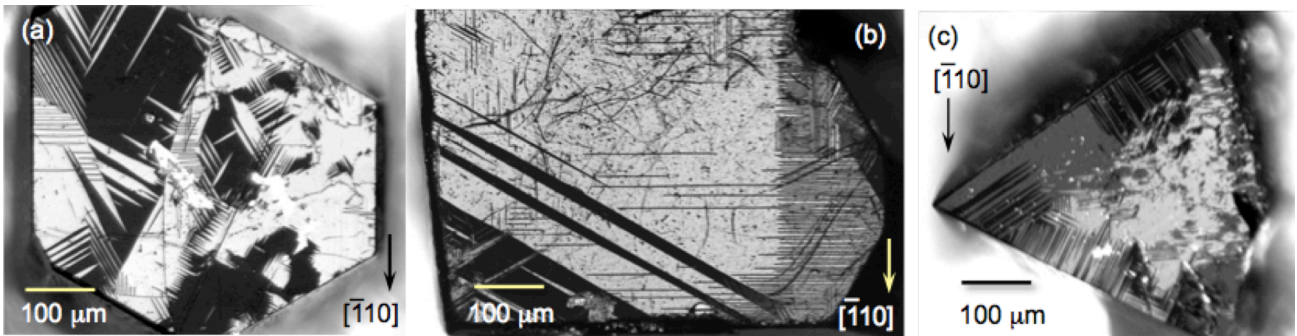

**Fig. 1.** (Color online) Photographs of crystals taken at 10 K using a polarizing microscope in a crossed-Nicols setup with a horizontal polarizer approximately along [11–2] and a vertical analyzer along [–110]. Three $Cd_2Re_2O_7$ crystals with RRR = 150 (a), 30 (b), and 5 (c) are shown. The crystal surface observed is a pristine (111) plane for each case. The photographic contrast has been modified so as to make the domain pattern clearly visible.

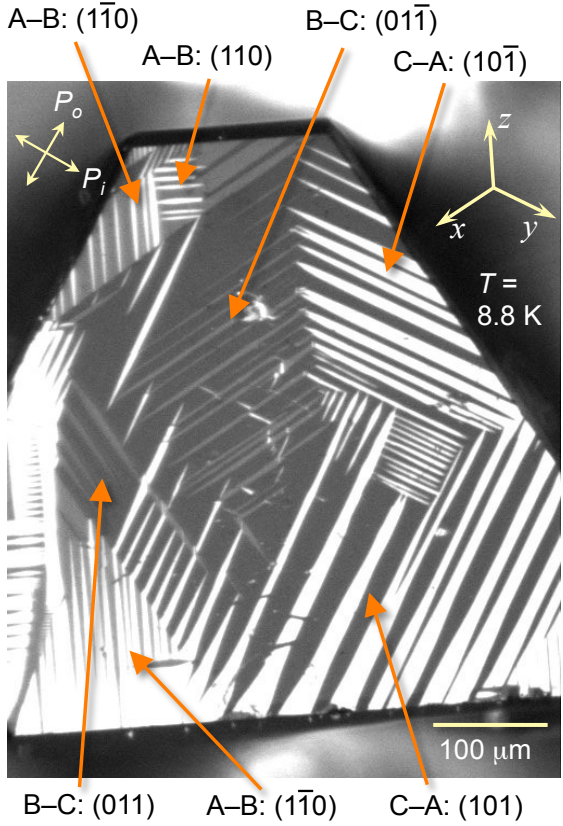

**Fig. 2.** (Color online) Typical twin-domain image of a $Cd_2Re_2O_7$ crystal with RRR = 60, which was taken at 8.8 K on a pristine (111) surface using a polarizing microscope with a polarizer approximately along [−12−1] ($P_i$) and an analyzer along [10−1] ($P_o$). The contrast of the image has not been adjusted. The identification of domains and their interfaces are based on the corresponding schematic drawing in Fig. 3(a).

We focus on the crystal with the best contrast shown in Fig. 2 from now on. The crystal surface is divided into several regions with bands of 10–50 μm width. Three levels of interference color are distinguished: dark, gray, and bright. Possible domains and their twins are schematically depicted in Fig. 3(a). Domains with their tetragonal $c$ axis aligned along the $x$, $y$, and $z$ axes are named domains A, B, and C, respectively. Between two domains, two sets of twin interfaces are generated, {1−10} and {110}, which are perpendicular to and inclined by 54.7° from the (111) plane, respectively. For example, twins with the (1−10) interface, A–B:(1−10) [Fig. 3(b)], and with the (110) interface, A–B:(110) [Fig. 3(c)] are generated between domains A and B. Note that their interfaces appear as sets of parallel lines orthogonal to each other on the (111) plane. As a result, six types of twin domains can be formed and observed on the (111) crystal surface.

Referring to Fig. 2, all six twin domains are uniquely identified from the orientations of the bands. For example, the dark–bright bands in the bottom right region correspond to the C–A:(101) twin domains, and the gray–bright bands at the bottom left are the A–B:(1−10) twin domains. This means that the common bright bands are the A domains. Similarly, the gray and dark bands are uniquely determined to be domains B and C, respectively.

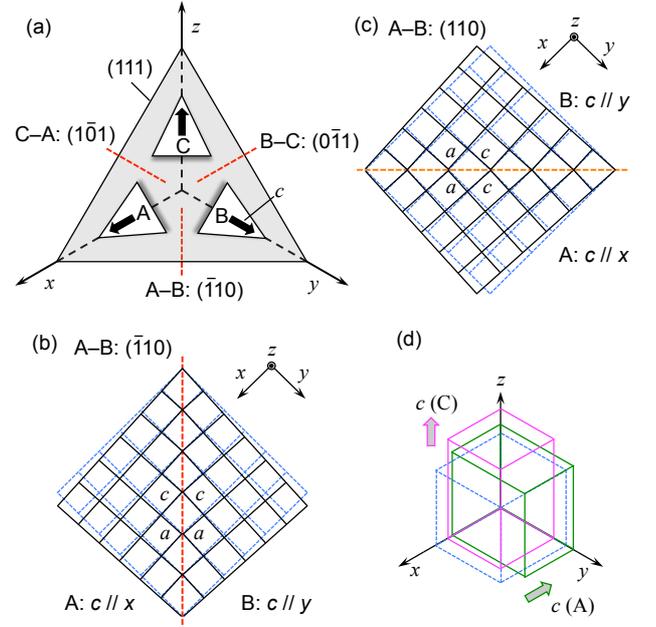

**Fig. 3.** (Color online) (a) (111) crystal surface illustrating the alignment of three possible domains with the tetragonal $c$ axis oriented along $x$ (domain A), $y$ (B), and $z$ (C) in the cubic notation. One set of twin interfaces perpendicular to the (111) plane are shown by the red broken lines, such as (1−10) for the A–B twin. (b) Lattice deformation at the A–B:(1−10) twin viewed along $z$ for $c > a$. The blue dashed square lattice represents the corresponding original cubic lattice of phase I. The tetragonal distortion is exaggerated for clarity. (c) Another twin domain between domains A and B, whose interface is not perpendicular to the (111) plane, C–A:(110), viewed along $z$. (d) Possible tetragonal deformations for phases II and III. Domain C with the elongated $c$ axis along $z$ for phase II (blue tetragonal prism) may be transformed at $T_{s2}$ to domain A with the contracted $c$ axis along $x$ (or $y$) for phase III (green tetragonal prism).

Next, we describe the temperature evolution of the twin domains. Figure 4 shows a series of polarizing microscope images taken with decreasing temperature for the crystal shown in Fig. 2 in a different temperature run. Without contrast adjustment, the variation of the brightness indicates the development of optical anisotropy. The dark uniform image at 215 K, above $T_{s1}$, is from the isotropic cubic phase I. At 180 K, just below $T_{s1}$ in phase II, two areas with bands are discerned, which correspond to the B–C:(0−11) and A–B:(1−10) twins. The image is brighter at 130 K with another area appearing at the bottom left, which is the C–A:(101) twin. Notably at 115 K, just below $T_{s2}$, a dramatic change occurs in the twin pattern, demonstrating the presence of a well-defined phase transition to phase III. Upon further cooling to 80 K, the contrast is enhanced, and the crystal is eventually divided into six twin domains; in addition to lamellar domains, lenslike domains are observed at the bottom. The domain pattern is then largely maintained down to 8.4 K.

Let us compare the two images shown in Figs. 2 and 4 at the lowest temperatures, which were recorded from the same crystal in different temperature cycles. They resemble each other, in particular, the positions of the boundaries between different twin domains are almost the same. However, it is noticed by careful examination that the A–B:(110) twin at the top left of Fig. 2 flips to the A–B:(1−10) twin in Fig. 4(a) (8.4 K). A reverse flip is also observed between the two images at the bottom left. Thus, the interface flips while keeping the same combination of domains, which may be more energetically favorable than changing the direction of the $c$ axis in one domain. Anyway, it is concluded that the distribution of twin domains is mostly reproduced in different temperature cycles, which may be determined by immobile crystalline defects.

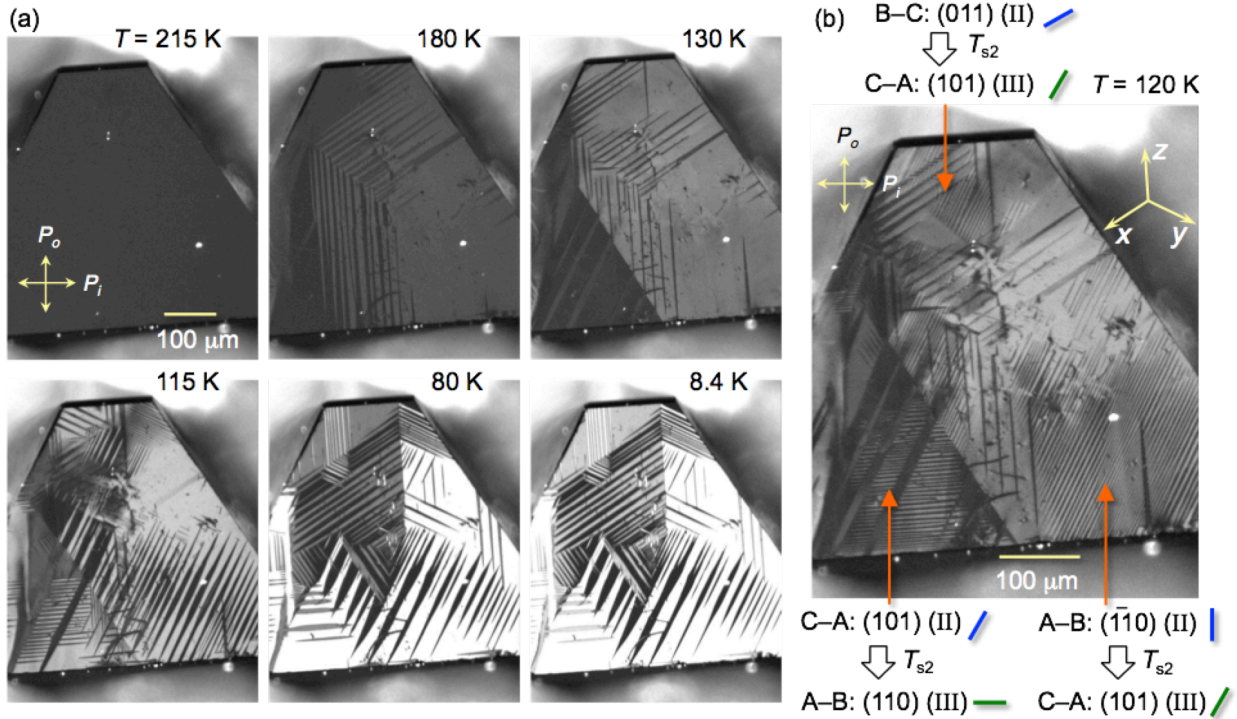

**Fig. 4.** (Color online) (a) Temperature evolution of the polarizing microscope images of the $Cd_2Re_2O_7$ crystal shown in Fig. 2 at $T$ = 8.4–215 K in a crossed-Nicols setup. (b) Image taken at 120 K, close to $T_{s2}$, showing the coexistence of high- and low-temperature twin domains. The changes in the twins from phase II to III upon cooling in three regions are described. The blue or green bar on the right of each domain description indicates the direction of the twin habit in the image.

Figure 4(b) is an enlarged image taken at 120 K, close to $T_{s2}$. In addition to the domain pattern observed at 130 K, fine stripes with ~5 μm intervals appear, which correspond to the pattern appearing at 115 K. For example, at the bottom right, the A–B:(1–10) twin from phase II and the C–A:(101) twin from phase III overlap. This coexistence indicates that the $T_{s2}$ transition is of the first order, in good agreement with the previous results.[2]

The fact that the domain pattern changes dramatically at $T_{s2}$ suggests that the $c$ axis flips by 90° at each domain across the transition; if there was no change in the direction of the $c$ axis, one would expect no change in the domain pattern. For example, the above-mentioned change from the A–B:(1–10) twin in phase II to the C–A:(101) twin in phase III must be caused by switching of domains, A –> C and B –> A, by flipping the $c$ axis in each domain. There is another possibility that the switching B –> C occurs while keeping the A domain. However, this may be unfavorable in terms of strain energy.

The previous discussion based on the Landau theory reveals that the tetragonality $t$, defined as $t = c/\sqrt{2}a$ for the body-centered-tetragonal cell, should change at $T_{s2}$: $t > 1$ for phase II and $t < 1$ for phase III, or vice versa,[13] as addressed in detail later. If this is the case, flipping of the $c$ axis across $T_{s2}$ is naturally expected, as depicted in Fig. 3(d), because compressing (elongating) the already elongated (compressed) $c$ axis in phase II at the transition to phase III must cause a larger strain than exchanging the $c$ axis with the intermediate $a$ axis.

### 3.2 Effects of magnetic fields on domain formation

The cubic-to-tetragonal transition at $T_{s1}$ causes not only crystalline anisotropy but also magnetic anisotropy. When a magnetic field is applied to a crystal, the magnetic energy changes as $MH = \chi H^2$; thus, the distribution of domains must be influenced by the anisotropy of the magnetic susceptibility $\chi$. In particular, for $Cd_2Re_2O_7$ with minimal lattice distortion, the magnetic energy may become sufficiently relatively large to align domains along the magnetic field and thus enable domain control by a magnetic field.

Figure 5 shows three sets of magnetic susceptibility $\chi$ data obtained from a crystal with RRR = 30 at magnetic fields along the [001], [110], and [111] directions. To maximize the effects, a large field of 7 T was applied to the crystal. Each measurement was performed first upon heating from 2 K after cooling in a zero field (ZFC) and then upon cooling in the same field (FC) at a rate of approximately 1 K/min. As reported previously, $\chi$ markedly decreases below $T_{s1}$, which is now found to be irrespective of the field direction; this decrease has been attributed to a reduction in the density of states associated with Fermi surface splitting due to inversion-symmetry breaking at $T_{s1}$.[2] The transition temperature, defined as an inflection point in the graph of $\chi$, is 204 K, slightly higher than previously reported.

Below $T_{s1}$, there is a clear difference between the ZFC and FC curves for $H \parallel [001]$: the decrease is larger for the ZFC curve. Then, at $T_{s2}$, both curves start to increase, the magnitude of the increase being larger for the FC curve. On the other hand, for $H \parallel [110]$, the ZFC and FC curves deviate from each other similarly to those for $H \parallel [001]$ below $T_{s1}$, while at $T_{s2}$, they change in opposite directions: the ZFC curve increases and the FC curve decreases; thus, the difference between them is reduced at lower temperatures. In sharp contrast, for $H \parallel [111]$, the ZFC and FC curves almost overlap each other with small humps at around $T_{s2}$. It is revealed from the data for $H \parallel [001]$ and $[110]$ that the $T_{s2}$ transition takes place in a temperature window of 116–112 K. Defining $T_{s2}$ as the midpoint, it is determined to be 114 K, which must correspond to the intrinsic transition temperature, as discussed in the examination of the sample quality.[2)]

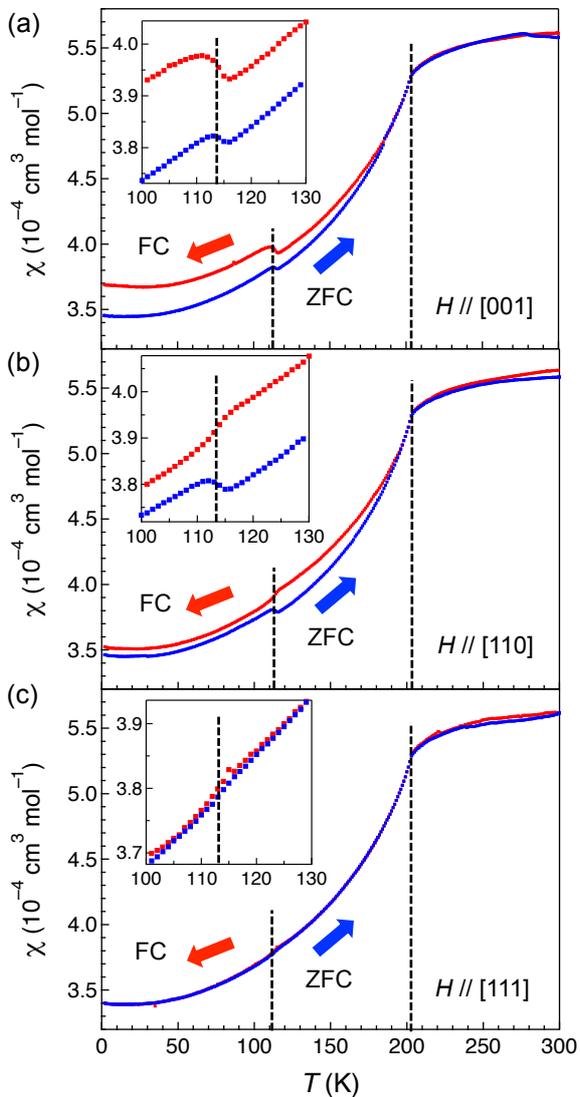

**Fig. 5.** (Color online) Magnetic susceptibilities of a $Cd_2Re_2O_7$ crystal (RRR = 30) in a magnetic field of 7 T parallel to the [001] (a), [110] (b), and [111] (c) directions. Each measurement was first carried out in a magnetic field of 7 T upon heating to 300 K after the crystal was cooled to 2 K in a zero field (ZFC) and then upon cooling in the same field (FC).

The three FC curves from Fig. 5 are compared in Fig. 6. In phase II, the [001] and [110] curves overlap and the [111] curve is lower, while $\chi$ decreases in the order of $H \parallel [001]$, [110], and [111] in phase III. These clear differences and the appearance of thermal "hysteresis" between the ZFC and FC curves below $T_{s1}$ strongly indicate that the alignment of domains actually takes place upon applying magnetic fields. Note that these significant field effects have been observed for a few crystals, but the magnitudes and hystereses vary from crystal to crystal; the crystal with the moderate RRR of 30 used for the $\chi$ measurements in Fig. 5 exhibited the most apparent field effects. The domain distribution may be critically influenced by certain crystalline defects, which cannot always be estimated from the RRR.

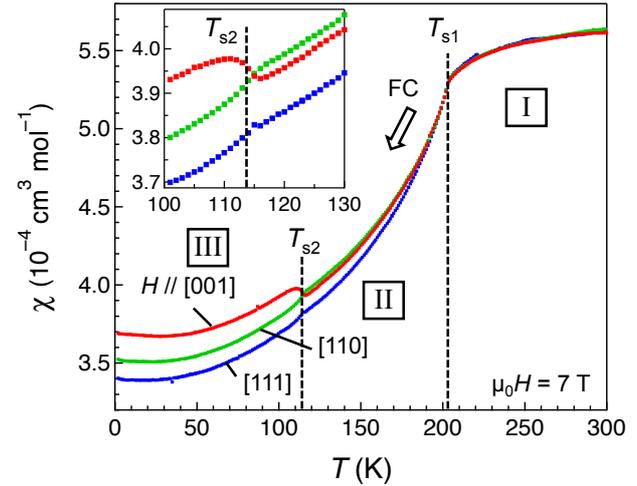

**Fig. 6.** (Colour online) Comparison of the magnetic susceptibility recorded in the FC process in a magnetic field of 7 T parallel to the [001], [110], and [111] directions from Fig. 5. The data for $H \parallel [001]$ and [110] have been multiplied by factors of 1.0399 and 1.0368, respectively, so that they coincide with the data for $H \parallel [111]$ at 200 K, taking into account the small deviations caused by differences in the crystal shape along the fields and the experimental setup.

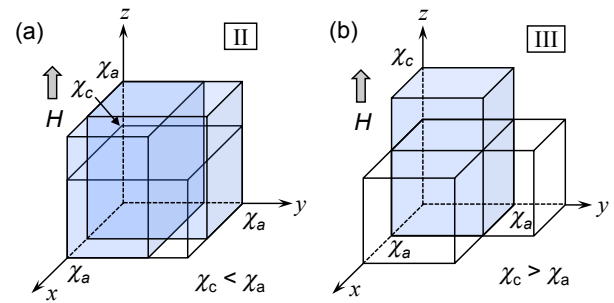

**Fig. 7.** (Color online) Schematic representations of the selection of domains for phases II (a) and III (b). When a magnetic field of sufficient magnitude is applied along the $z$ axis, either of the two shaded domains with $c \parallel x$ or $y$ is selected for phase II ($\chi_c < \chi_a$) and one shaded domain with $c \parallel z$ is selected for phase III ($\chi_c > \chi_a$).

In general, the magnetic anisotropy in a tetragonal crystal has two possibilities: $\chi_c$ is smaller or larger than $\chi_a$. As schematically depicted in Fig. 7, for $H \parallel [001]$, for a smaller $\chi_c$ two domains, A ($c \parallel x$) and B ($\parallel y$), are selected, while only domain C ($\parallel z$) is chosen for a larger $\chi_c$. Then, the resulting $\chi_{av}$ is $\chi_a$ and $\chi_c$ for $\chi_c < \chi_a$ and $\chi_c > \chi_a$, respectively, as summarized in Table I. On the other hand,

5 / 8

**Table I.** Magnetic susceptibility $\chi_{av}$ expected in the case of perfect domain alignment with the three directions of the magnetic field for phases II and III of $Cd_2Re_2O_7$. Provided that $\chi_c < \chi_a$ for phase II and vice versa for phase III, the type of domains and the relative magnitude of $\chi_{av}$ are uniquely decided; L: large, M: medium, S: small. The tetragonality $t$, defined as $t = c/\sqrt{2}a$ for the body-centered-tetragonal cell, may be larger and smaller than 1 for phases II and III, respectively, or vice versa.

| Phase | $H$ | Domain | $\chi_{av}$ | Magnitude |
|---|---|---|---|---|
| II | [001] | double | $\chi_a$ | L |
| $\chi_c < \chi_a$ | [110] | single | $\chi_a$ | L |
| ($t > 1$) | [111] | triple | $(2\chi_a + \chi_c)/3$ | S |
| III | [001] | single | $\chi_c$ | L |
| $\chi_c > \chi_a$ | [110] | double | $(\chi_a + \chi_c)/2$ | M |
| ($t < 1$) | [111] | triple | $(2\chi_a + \chi_c)/3$ | S |

domains C and A + B are selected for $H \parallel [110]$ in the two cases, respectively, and three domains always coexist for $H \parallel [111]$. Accordingly, the different $\chi_{av}$ expected for complete domain alignment are calculated as in Table I. Thus, merely from the magnitude relation between $\chi_c$ and $\chi_a$, the relative magnitudes are obtained as $\chi[001] = \chi[110] > \chi[111]$ for $\chi_c < \chi_a$ and $\chi[001] > \chi[110] > \chi[111]$ for $\chi_c > \chi_a$. These relative magnitudes exactly match what we have observed for phases II and III, respectively, indicating that switching of the anisotropy in $\chi$ indeed occurs at $T_{s2}$. It is also consistent with the lack of anisotropy for $H \parallel [111]$, in which the three domains are always equivalent.

The observed thermal hystereses for $H \parallel [001]$ and [110] below $T_{s1}$ are explained as follows. Domain alignment by a magnetic field on randomly frozen domains produced in ZFC is not possible at low temperatures. In contrast, in a slow FC process from above $T_{s1}$, only domains with their easy axes toward the field tend to be created, which are domains A and B for $H \parallel [001]$ and domain C for $H \parallel [110]$; thus, the FC curve shifts upward to above the ZFC curve, as observed. On the other hand, the dramatic changes in $\chi$ at $T_{s2}$ are due to the switching of the easy axis: $\chi_c < \chi_a$ in phase II and $\chi_c > \chi_a$ in phase III; domains A and B become domain C for $H \parallel [001]$, and domain C becomes domains A and B for $H \parallel [110]$. The observed changes at $T_{s2}$ are explained qualitatively by the anisotropy switching of $\chi$.

Let us move to a quantitative estimation of the domain alignment. The experimental magnetic susceptibilities $\chi_{meas}$ at 2 K in Fig. 6 are $3.69 \times 10^{-4}$, $3.52 \times 10^{-4}$, and $3.40 \times 10^{-4}$ cm$^3$ mol$^{-1}$ for $H \parallel [001]$, [110] and [111], respectively, which are compared with the values of $\chi_{av}$ given in Table I expected for complete domain alignment for phase III. Then, $\chi_c$ is uniquely determined to be $\chi[001] = 3.69 \times 10^{-4}$ cm$^3$ mol$^{-1}$, and $\chi_a$ is calculated to be $3.35 \times 10^{-4}$ cm$^3$ mol$^{-1}$ from $\chi[110] = (\chi_a + \chi_c)/2$ or $3.26 \times 10^{-4}$ cm$^3$ mol$^{-1}$ from $\chi[111] = (2\chi_a + \chi_c)/2$. Thus, the three experimental $\chi_{meas}$ values are reasonably reproduced with $\chi_c \sim 3.7 \times 10^{-4}$ cm$^3$ mol$^{-1}$ and $\chi_a \sim 3.3 \times 10^{-4}$ cm$^3$ mol$^{-1}$ for phase III, indicating that nearly perfect domain alignment has been achieved. The anisotropy thus obtained is 11% ($\chi_c > \chi_a$) in phase III.

For phase II, we take $\chi_{meas}$ at 120 K: $3.96 \times 10^{-4}$, $3.99 \times 10^{-4}$, and $3.86 \times 10^{-4}$ cm$^3$ mol$^{-1}$ for $H \parallel [001]$, [110], and [111], respectively. Also assuming complete alignment would give $\chi_c \sim 3.6 \times 10^{-4}$ cm$^3$ mol$^{-1}$ and $\chi_a \sim 4.0 \times 10^{-4}$ cm$^3$ mol$^{-1}$, the anisotropy is 10% ($\chi_c < \chi_a$) in phase II. Interestingly, the anisotropies in $\chi$ are nearly equal and the direction is switched at $T_{s2}$. These large magnetic anisotropies in the nonmagnetic metal must be caused by the spin-split Fermi surfaces.

## 4. Discussion

### 4.1 Comparison with the Landau theory

The successive structural transitions of $Cd_2Re_2O_7$ have been discussed in terms of the Landau theory.[13-15] The adapted order parameters ($\eta_1$, $\eta_2$), which span the $E_u$ representation of the cubic point group $O_h$, are linear combinations of the displacements of the four Re atoms of one tetrahedron from their ideal positions in phase I, ($x_m$, $y_m$, $z_m$) with $m = 1, 2, 3, 4$:

$$\eta_1 = (X - Y)/\sqrt{2}, \eta_2 = (X + Y - 2Z)/\sqrt{6}, \quad (1)$$

where $X = (x_1 + x_2 - x_3 - x_4)/2$, $Y = (y_1 - y_2 + y_3 - y_4)/2$, and $Z = (z_1 - z_2 - z_3 + z_4)/2$. Phases I, II, and III are characterized by ($\eta_1$, $\eta_2$) = (0, 0), (0, $\eta_2$), and ($\eta_1$, 0), respectively.[14]

By choosing appropriate coefficients for the terms in the thermodynamic potential, the second-order transition from phase I ($Fd\bar{3}m$) to phase II ($I\bar{4}m2$) and the subsequent first-order transition to phase III ($I4_122$) are reproduced;[14] here we take the $E_u$ order parameter, which gives a complete explanation of our observations as mentioned below, although another order parameter of $T_{2u} + T_{1g}$ has been suggested on the basis of second-harmonic optical anisotropy measurements, which requires a crystal structure of space group $I\bar{4}2d$ or its subgroup $I\bar{4}$ for phase II.[3-5] Note that although these transitions involve structural changes, the true driving force must be electronic, which is ascribed to the Fermi liquid instability of the SOCM.[1] Considering the metallic nature and the point group symmetry, the two low-temperature phases are called "piezoelectric metals".[13]

For the structural deformations at the two transitions, the contribution of the strain components ($e_1$, $e_2$, $e_3$) to the thermodynamic potential is considered.[13] They are given as

$$e_1 = e_2 = -(\lambda_1 + \lambda_2)\eta_2^2; e_3 = -(\lambda_1 - 2\lambda_2)\eta_2^2 \quad (2)$$

for phase II and as

$$e_1 = e_2 = -(\lambda_1 - \lambda_2)\eta_1^2; \quad e_3 = -(\lambda_1 + 2\lambda_2)\eta_1^2 \quad (3)$$

for phase III, where $\lambda_1$ and $\lambda_2$ are parameters containing the elastic stiffness constants. The experimental finding that the difference between $e_1$ and $e_3$, that is, the tetragonality, is always very small, strongly indicates that $\lambda_1 < 0$ and $|\lambda_1| \gg |\lambda_2|$.[13] Importantly, $\lambda_2$ determines the tetragonal deformation, which should change between phases II and III. In the case of $\lambda_2 > 0$, $e_1 < e_3$ ($t > 1$) for phase II and $e_1 > e_3$ ($t < 1$) for phase III [Fig. 3(d)] and vice versa for $\lambda_2 < 0$. This is consistent with the observed dramatic changes in the twin-domain patterns at $T_{s2}$, suggesting that the $c$ axis of the tetragonal domain actually flips.

In the previous structural studies, the corresponding change in the tetragonality has not been observed: it was always found that $t < 1$ below $T_{s1}$ down to ~15 K.[7,16] However, because of the small tetragonality, one has to be careful when concluding this experimentally. Our preliminary high-resolution neutron scattering experiments showed that the switching of $t$ indeed takes place at $T_{s2}$. We plan to perform another experiment to confirm this finding.

On the other hand, the Landau theory can also be applied to the magnetic response.[15] In an applied field **H**, the magnetic part of the free energy is

$$F_M = A\mathbf{M}^2 - \mathbf{MH} + \gamma(\eta_1^2 + \eta_2^2)\mathbf{M}^2 + \delta[(\eta_1^2 - \eta_2^2)(2M_z^2 - M_x^2 - M_y^2) + 2\sqrt{3}\eta_1\eta_2(M_x^2 - M_y^2)], \quad (4)$$

where **M** is the magnetization and the product $\eta_1\eta_2$ is zero in all three phases. The inverse magnetic susceptibility is calculated to be

$$\chi_{zz}^{-1} = \chi_0^{-1} + 2\gamma\eta_2^2 - 4\delta\eta_2^2 \quad (5)$$
$$\chi_{xx}^{-1} = \chi_0^{-1} + 2\gamma\eta_2^2 + 2\delta\eta_2^2 \quad (6)$$

for phase II and

$$\chi_{zz}^{-1} = \chi_0^{-1} + 2\gamma\eta_1^2 + 4\delta\eta_1^2 \quad (7)$$
$$\chi_{xx}^{-1} = \chi_0^{-1} + 2\gamma\eta_1^2 - 2\delta\eta_1^2 \quad (8)$$

for phase III, where $\chi_0^{-1}$ is the susceptibility of phase I. Since $\chi$ decreases markedly below $T_{s1}$, $\gamma$ must be positive and relatively large.[15] The anisotropy arises from the third term. The present observation that $\chi_c < \chi_a$ for phase II and $\chi_c > \chi_a$ for phase III is perfectly consistent with $\delta < 0$; the observed anisotropies of ~10% indicate that $\delta$ is much smaller than $\gamma$. Therefore, our experimental observations on the tetragonal distortion and magnetic susceptibility are qualitatively interpreted by the Landau theory for the $E_u$ order parameter.

### 4.2 Domain control

In order to investigate possible odd-parity multipole orders of $Cd_2Re_2O_7$, it is necessary to study the anisotropies in transport properties and magnetoelectric or Edelstein effects in the low-temperature phases. For this, understanding of domain formation and distribution is indispensable. First, we consider domain control by a magnetic field. It is likely that the anisotropies of ~10% in $\chi$ dominate domain formation by suppressing twin domains introduced to reduce the elastic energy arising from the tetragonality of $\leq 0.05\%$, which is in fact evidenced in the present study at a magnetic field of 7 T.

The change in the twin domain pattern observed by polarizing microscopy must be affected in the presence of a magnetic field. For example, the change from the A–B twin to the C–A twin at $T_{s2}$ [Fig. 4(b)] may be modified so as to remove the twin interface for $H \parallel [001]$: both domains A and B become domain C. Since domain C in phase II is not affected, a single-domain crystal of phase III would be obtained for $H \parallel [001]$, as depicted in Fig. 7, if the magnetic energy surpassed the elastic energy and the energy required to make the domain walls mobile. On the other hand, one expects a single-domain crystal of phase II for $H \parallel [110]$.

Note, however, that pinning by crystalline defects may impede the domain-wall motion and prevent complete domain alignment. In particular, defects that pin boundaries between twin domains with different orientations, such as those observed in Figs. 1 and 2, must be critical. On the other hand, this means that, for a cleaner crystal, complete domain alignment will be achieved by applying a lower magnetic field or even at low temperatures after ZFC.

On the other hand, domain control by mechanical strain may also be possible. In a preliminary experiment, we observed a slight change in the twin domain pattern by polarizing microscopy when a tensile stress was applied to a crystal placed on a piezoelectric device along [1–10] in the (111) plane at 8.4 K. The largest strain effect is expected for a [001] strain. If $\lambda_2 > 0$ ($t > 1$ for phase II and $t < 1$ for phase III), a tensile stress along [001] would stabilize domain C for phase II and domains B and C for phase III. Therefore, a single-domain crystal could be obtained by applying a tensile stress for phase II and a compressive stress for phase III. Such experiments are in progress. In addition, in combination with magnetic field effects, a compressive strain and a magnetic field, both along [001], should achieve a perfect single-domain crystal of phase III.

### 4.3 "Invisible" domains

Finally, note that there must be two types of "invisible" domains in every twin lamella that are not observable by polarizing microscopy for either of the low-temperature phases of $Cd_2Re_2O_7$. In phase II, these domains are connected by antiphase boundaries originating from the inversion-symmetry breaking, whereas in phase III, they are two chirality domains associated with the fourfold screw axis. In fact, Harter et al. observed two regions with different signals in one twin lamella for phase II in their second-harmonic optical anisotropy measurements, which they claimed to be domains of an odd-parity nematic order.[3] It is likely that the invisible domains of structural origin and the electronic multipolar domains are identical to each other, because the electronic instability of the SOCM may couple with the lattice degree of freedom. In a single-domain crystal attained by applying an appropriate magnetic field or mechanical stress, these higher-order domains could be controlled and selected by the magnetoelectric or Edelstein effects expected for a spin-split Fermi surface in an odd-parity multipole phase,[17] which will be experimentally examined in future work.

## 5. Conclusion

Regarding the successive phase transitions of $Cd_2Re_2O_7$, polarizing microscopy observations of a (111) crystal surface revealed the formation of twin domains upon cooling below $T_{s1} \sim 200$ K and a dramatic change in the twinning pattern at $T_{s2} \sim 120$ K, which suggests a reversal of tetragonality. Magnetic susceptibility measurements revealed significant domain alignment when applying a magnetic field of 7 T upon cooling, which is due to the anisotropy of ~10% in the magnetic susceptibility of the low-temperature phases. The anisotropy is reversed at $T_{s2}$: $\chi_c < \chi_a$ above $T_{s2}$ and vice versa below $T_{s2}$, which causes reorientation of the tetragonal domains at $T_{s2}$.

**Acknowledgments** We thank J. Yamaura, M. Takigawa, T. Arima, Y. Motome, and M. Iwata for helpful comments and discussion. Y.M. was supported by the Materials Education Program for the Future Leaders in Research, Industry, and Technology (MERIT) created by the Ministry of Education, Culture, Sports, Science and Technology of Japan (MEXT). This work was partially supported by KAKENHI Grant No. 18H01169 and the Core-to-Core Program for Advanced Research Networks given by the Japan Society for the Promotion of Science (JSPS).